\documentclass[prl,aps,twocolumn,showpacs]{revtex4}

\usepackage{graphics}
\usepackage{graphicx}
\usepackage{amssymb,amsmath}
\usepackage{bm}

\newcommand{\mytitle}[1]{

 \twocolumn[\hsize\textwidth\columnwidth\hsize

 \csname@twocolumnfalse\endcsname #1 \vspace{1mm}]}

\newcommand{\beq}{\begin{equation}}
\newcommand{\eeq}{\end{equation}}
\newcommand{\bea}{\begin{eqnarray}}
\newcommand{\eea}{\end{eqnarray}}

\begin{document}

\title{Quantum criticality out of equilibrium in the pseudogap Kondo model}
\author{ Chung-Hou Chung and Kenneth Yi-Jie Zhang}
\affiliation{
Electrophysics Department, National Chiao-Tung University,
HsinChu, Taiwan, R.O.C. }
\date{\today}

\begin{abstract}
We theoretically investigate the non-equilibrium quantum 
phase transition in a generic setup: the pseudogap Kondo model 
where a quantum dot couples to two--left (L) and right (R)--voltage-biased 
fermionic leads with power-law density of states (DOS) with respect to their 
Fermi levels $\mu_{L/R}$, 
$\rho_{c, L(R)}(\omega)\propto |\omega-\mu_{L(R)}|^{r}$, and $0<r<1$. 
In equilibrium (zero bias voltage) and for $0<r<1/2$, with increasing Kondo 
correlations, in the presence of particle-hole  
symmetry this model exhibits a quantum phase transition from a unscreened 
local moment (LM) phase to the Kondo phase. Via a controlled frequency-dependent
renormalization group (RG) approach, we compute analytically and numerically 
the non-equilibrium conductance, conduction electron T-matrix and 
local spin susceptibility at finite bias voltages near criticality. 
The current-induced decoherence shows distinct nonequilibrium scaling,  
leading to new universal non-equilibrium quantum 
critical behaviors in the above observables. 
Relevance of our results for the experiments 
is discussed. 
\end{abstract}

\maketitle


 
\emph{Introduction.} 
Quantum phase transitions (QPTs)\cite{subir}, 
the continuous phase transitions occur at zero temperature 
due to quantum fluctuations, in strongly correlated electron systems 
have attracted much attention over the last three decades. Near 
the quantum critical points (QCPs) associated with QPTs, 
thermodynamic properties exhibit non-Fermi liquid properties and 
universal scalings. 
Recently, due to high tunability, 
nano-devices, such as: quantum dots in the Kondo regime\cite{Goldhaber,Hewson}, 
offer a new opportunity to study QPTs. 
In particular, understanding QPTs in nano-systems under nonequilibrium 
conditions has become one of the 
outstanding emergent subjects in condensed matter 
physics with great fundamental importance\cite{NoneqQPT,chung,chung2}.  
In Ref.~\cite{chung}, the authors discovered the distinct non-equilibrium profile in 
transport near the 
localized-delocalized QPT of the Kosterlitz-Thouless (KT) type  
in a generic voltage-biased dissipative resonance-level (quantum dot) from 
its equilibrium properties at finite temperatures. 
The current-induced decoherence rate 
smearing out the transition shows highly non-linear voltage 
dependence, resulting in these distinct behaviors near QPT.\\ 
In this paper, we investigate the non-equilibrium quantum criticality 
in a different class of generic nano-setup--the pseudogap Kondo (PGK) 
model\cite{fradkin,GBI,insi,lars,florens} in a quantum dot\cite{hurpsg}. 
We consider a Kondo quantum dot couples to two--left (L) 
and right (R)--fermionic 
leads with a power-law (pseudogap) density-of-states (DOS) 
which vanishes at the Fermi level $\mu_{L(R)}=\pm V/2$, 
$\rho_{c, L(R)}(\omega)\propto |\omega-\mu_{L(R)}|^{r}$ 
with $0<r<1$. Possible realizations of the pseudogap leads 
include: $d-$wave superconductors ($r=1$)\cite{lars}, 
graphene\cite{vojtagraphene} 
($r=1$), one-dimensional Luttinger systems 
($r>0$)\cite{GBI}, and quantum dots embedded in a Aharonov-Bohm ring 
($r=2$)\cite{nancy}. 
In equilibrium ($V=0$) and for $0<r<1/2$, with decreasing the Kondo 
couplings the particle-hole (p-h) symmetric PGK model exhibits a ``true'' 
QPT (distinct from QPT of 
the KT type\cite{trueQPT}) from the Kondo screened phase 
to the unscreened local moment (LM) phase\cite{GBI,lars}. 
Near QCP separating these two phases, 
all observables in equilibrium exhibit universal power-law 
scalings and have been extensively studied\cite{lars,florens}. 
Nevertheless, there is still lack of understanding regarding 
their corresponding out-of-equilibrium quantum critical properties.   
We shall address below this issue with a focus on the 
universal nonequilibrium scaling behaviors near QCP. \\
\emph{The model and the RG approach.}
The Hamiltonian of the particle-hole (p-h) symmetric PGK model reads:
\begin{equation}
H=\sum_{k\alpha} (\epsilon_{k\alpha}-\mu_{\alpha}) c^\dag_k c_k 
+ \sum_{\alpha ,\alpha ^{\prime },k,k^{\prime },\sigma ,\sigma
^{\prime }} J_{\alpha ,\alpha ^{\prime }} 
\mathbf{S}^{dot}\cdot \mathbf{S}^e_{\alpha'\alpha}
\label{H-Kondo}
\end{equation}
where $\mathbf{S}^{dot}=f^\dag_{\sigma'} 
\tau_{\sigma'\sigma} f_{\sigma}$, $\mathbf{S}^e_{\alpha\alpha'}= c_{\alpha ^{\prime },k^{\prime
},\sigma ^{\prime }}^{\dagger }\tau _{\sigma ^{\prime }\sigma }c_{\alpha
,k,\sigma }$ are the spin-1/2 operators of the electron 
on the dot and in the leads, respectively, $\tau$ are Pauli matrices,  
and $\alpha, \alpha' =L/R$,  
$\sigma,\sigma'=\uparrow\downarrow$ are the 
lead and spin indices, respectively.  
$c^{\dagger}_{\alpha, k, \sigma}$ is the electron 
creation operator for the lead $\alpha$ with Fermi energies being 
$\mu_{L/R}=\pm V/2$, and $f_{\sigma}$ is the pseudofermion operator. 
The conduction electron leads show power-law (pseudogap) DOS 
with respect to their Fermi levels $\mu_{L/R}$, 
$\rho_{c, L(R)}(\omega)\propto |\omega-\mu_{L(R)}|^{r}$, and $0<r<1$.     
In the Kondo regime, the single-occupancy constraint of the 
pseudo-fermions is imposed: 
$\sum_{\sigma }f_{\sigma }^{\dagger }f_{\sigma }=1$. 
Here, the dimensionless inter-lead and intra-lead Kondo couplings 
are denoted by $g_{LR}= N_0 J_{LR}$, and 
$g_{LL}=g_{RR}=N_0 J_{LL}=N_0 J_{RR}$, respectively 
where $N_0 = \frac{1}{2 D_0}$ and $D_0$ is the bandwidth cutoff of the 
leads. For simplicity, we consider here the symmetrical Kondo 
couplings: $g_{\alpha\beta}=g$. 
In equilibrium, the one-loop RG scaling equation for $g$ reads 
$\frac{\partial g}{\partial \ln D} = r g - 2 g^2$\cite{lars}. 
The critical Kondo coupling 
$g_c=\frac{r}{2}$ separates the Kondo ($g>g_c$) from the unscreened 
local moment (LM) phase ($g<g_c$). Much of the equilibrium 
critical properties can be obtained from the cutoff dependence of the 
renormalized Kondo coupling: 
$g^{eq}(D) = \frac{g_c}{1+ (D/T^{\ast})^{-r}}$ with the crossover energy 
scale being 
$T^{\ast}= D_0 (\frac{|g_c-g_0|}{g_0})^{\frac{1}{r}}\propto |g_c-g|^{\nu}$ 
and the correlation length exponent being $\nu = 1/r$. 
At a finite bias voltage, however, the chemical potentials 
(Fermi levels) of the two leads are shifted by $\pm V/2$. 
Under various 
RG approaches, the Kondo interaction vertices in general depend not only 
on the cutoff scale $D$, but also on the electron energy 
(frequency)\cite{Rosch,fRG}.  
We employ here a weak coupling  
1-loop frequency-dependent RG approach of Ref.~\cite{Rosch,chung} 
which keeps track of energy of the incoming electrons. 
For $r=0$ our results 
agree excellently with those via a more sophisticated functional RG 
approach in Ref.\cite{woelflefRG}. 
Note that our weak coupling theory for the p-h symmetric model 
Eq.~\ref{H-Kondo} works well only for $r\rightarrow 0$. The QCP between LM 
and Kondo phases disappears for $r\ge 1/2$, and our theory 
breaks down for $r$ near $1$\cite{lars}. Note also that the  
above p-h symmetric QCP is stable 
against p-h asymmetry for $0<r<r^{\ast}=0.375$\cite{lars}. 
We therefore restrict ourselves to the p-h symmetric model 
for simplicity. The scaling equation for the Kondo couplings of our model 
under this approach reads\cite{Rosch,woelflefRG}: 
\begin{figure}[t]
\begin{center}
\includegraphics[width=6.0cm]{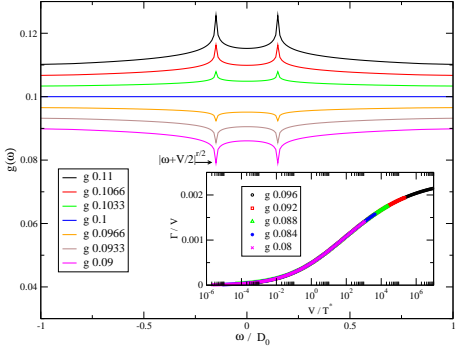}
\end{center}
\par
 \vskip -0.7cm
\caption{
(Color online)
Renormalized Kondo coupling $g(\protect\omega)$ for various bare 
couplings (in units of $D_0$) for $r=0.2$ ($g_c=0.1$). 
The bias voltage is $V = 0.3$. Inset: Universal scaling of $\Gamma/V$ as a 
function of $V/T^*$ with $\Gamma$ being the decoherence rate.}
\label{gpergzfigK}
\end{figure}
\begin{eqnarray}
\frac{\partial g(\omega)}{\partial \ln D}
&=&\sum_{\beta=-1,1} \tanh(\frac{D}{2T}) 
\left[ \frac{r}{2} g(\omega)-g^2(\omega) \right]\nonumber \\ 
& & \Theta \left( D- \left| \omega + \frac{\beta V}{2} + i\Gamma \right| \right)
\label{RGeq}\\
\Gamma &=&\pi \sum_{\alpha \alpha ^{\prime }}\int {d\omega f_{\omega }^{\alpha
}\left( 1-f_{\omega }^{\alpha ^{\prime }}\right) g^2(\omega )},  
\label{gamma}
\end{eqnarray}
where $\Gamma$ 
is the current-induced decoherence rate 
obtained from  the imaginary part of the pseudofermion 
self-energy\cite{Rosch}, 
$f_{\omega }^{\alpha}= \frac{1}{e^{\frac{\omega-\mu_{\alpha}}{T}}+1}$ is 
the Fermi function of the $\alpha$ lead and $k_B=\hbar=e=1$. Note that 
in equilibrium at a finite temperature $T$ 
the RG flows of the Kondo couplings are cut off by $T$; while as within the 
nonequilibrium RG approach they are cutoff by $\Gamma\ll V$, 
a much lower energy scale than $V$\cite{Rosch}. 
Distinct critical behaviors are therefore 
expected\cite{chung}. We shall focus below on what these distinct nonequilibrium quantum critical behaviors are.\\
We first solve Eq.~\ref{RGeq} and Eq.~\ref{gamma} for $g(\omega)$ 
self-consistently at $T=0$.  
As shown in Fig. 1, for $g> (<)g_c$, the renormalized Kondo couplings 
exhibit peaks (dips) at $\omega= \pm V/2$, indicating Kondo (local moment) 
phase; while $g(\omega)$ is completely flat at criticality $g=g_c$. 
The qualitative nature of these peaks (dips) in $g(\omega)$ agree 
well with Ref.~\cite{chung,chung2} as signatures of conducting (insulating) 
behavior. The height (depth) of the peaks (dips) get shorter (shallower) as 
one reaches to QCP from the Kondo (LM) phase. We restrict ourselves to 
the LM phase ($g\le g_c$) where the perturbative 
RG approach is controlled. 
The full analytical solution for $g(\omega)$ in the LM phase 
in the limit of $D\rightarrow 0$ is found to be: 
\begin{eqnarray}
g(\omega) &=& g + g_1(\omega) +g_2(\omega),\nonumber \\
g_1(\omega) &=&  \frac{ g_0  (1+\tilde{V}^{r})  
(|\tilde{\omega} - \frac{\tilde{V}}{2}|^{r} -1) }{2(1+ \tilde{V}^{r}) 
(1+|\tilde{\omega} - \frac{\tilde{V}}{2}|^{r})} 
\Theta(\tilde{D_0} - |\tilde{\omega} -\frac{\tilde{V}}{2}|) \nonumber \\
&+& \frac{g_c (\tilde{V}^{r}- |\tilde{\omega} - \frac{\tilde{V}}{2}|^{r})}
{2(1+ \tilde{V}^{r}) 
(1+|\tilde{\omega} - \frac{\tilde{V}}{2}|^{r})} 
\Theta(\tilde{V} - |\tilde{\omega} -\frac{\tilde{V}}{2}|)\nonumber \\
&+& (\omega \rightarrow -\omega),\nonumber \\
g_2(\omega) &=& 
(\frac{g_c \tilde{V}^{\frac{r}{2}}}{1+\tilde{V}^{r}}) 
\{ \frac{\tilde{V}^{\frac{r}{2}} -|\tilde{\omega}-
\frac{\tilde{V}}{2}|^{\frac{r}{2}} }
{1+ \tilde{V}^{\frac{r}{2}} 
|\tilde{\omega}-\frac{\tilde{V}}{2}|^{\frac{r}{2}}}\nonumber \\
&\times &
[\Theta(\tilde{\Gamma}- |\tilde{\omega}-\frac{\tilde{V}}{2}|) 
- \Theta(\tilde{D_0}-|\tilde{\omega}-\frac{\tilde{V}}{2}|)] \nonumber \\ 
&+& 
\frac{\tilde{\Gamma}^{\frac{r}{2}}-\tilde{V}^{\frac{r}{2}}}
{1+(\tilde{V}\tilde{\Gamma})^{\frac{r}{2}}} 
\Theta(\tilde{\Gamma}-|\tilde{\omega}-\frac{\tilde{V}}{2}|)\} \nonumber \\
&+& (\omega \rightarrow -\omega)
\label{gw-analytical}
\end{eqnarray}
with $\tilde{V}=\frac{V}{T^*}$, 
$\tilde{\omega}= \frac{\omega}{T^*}$, $\tilde{D_0}=\frac{D_0}{T^*}$, 
$\tilde{\Gamma}=\frac{\Gamma}{T^*}$, and $g_0$ being the bare Kondo coupling. 
The peaks (dips) of $g(\omega)$ 
near $\omega=\pm V/2$ shows a power-law behavior: $|g(\omega)-g(\omega=\pm V/2)| 
\propto |\omega \mp \frac{V}{2}|^{\frac{r}{2}}$ with a width of $\Gamma$. 
We furthermore 
find analytically via Eq.~\ref{gw-analytical} the universal scaling forms 
for $g(\omega=0, V)$, and $g(\omega=\pm V/2, V)$. These properties  
will be used in the following analysis to determine various novel 
nonequilibrium scaling behaviors in the LM phase:
\begin{equation}
g(\omega=0)=\frac{g_c}{1 + \left( \frac{V}{2T^*} \right)^{-r}},   
g(\omega=\frac{V}{2}) = \frac{g_c}{1 + \left( \frac{V\Gamma}{{T^*}^2} 
\right)^{-\frac{r}{2}}}. 
\label{g0gVover2}
\end{equation}
\begin{figure}[t]
\begin{center}
\includegraphics[width=7.25cm]{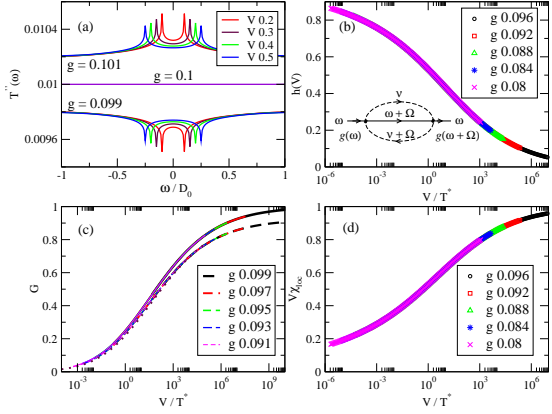}
\end{center}
\par
 \vskip -0.7cm
\caption{
(Color online)  (a). The imaginary part of the T-matrix $T^{"}(\omega)$ (in 
unit of $\frac{-3\pi}{8 N(0)}$) versus  
$V/T^*$ at $T=0$. (b). $h(V)$ defined 
in Eq. 9 versus $V/T^*$. Inset: the diagram for the T-matrix. 
(c).The $T=0$ nonequilibrium conductance $G(V)$ (solid lines) 
normalized to $\frac{3\pi g_c^2}{4}$ 
in the LM phase versus $V/T^*$ shows distinct scaling from the equilibrium counterpart, $G^{eq}(T\rightarrow V)$ (dashed lines). The dotted line is 
the analytical form via Eq. 11.  
(d). The scaling of 
$V\chi_{loc}$ versus $V/T^*$ with $\chi_{loc}$ being the local impurity susceptibility. The bare Kondo couplings in (a),(b),(c), and (d) are in unit of $D_0$, 
and $r=0.2$.}
\label{gpergzfigK}
\end{figure}
\emph{Nonequilibrium decoherence.} The current-induced 
decoherence $\Gamma$ which cuts off the RG flow 
is the key to understand  
nonequilibrium quantum criticality of the model as all nonequilibrium 
observables depend crucially on the scaling behavior of $\Gamma$. 
As shown in Fig.1 (Inset), $\Gamma/V$ in the LM phase 
exhibits perfect universal $V/T^*$ scaling over 
a wide range $10^{-6}<V/T^* <10^{6}$. 
We believe this slow crossover which extends  
over many decades is likely related to the large 
correlation length exponent $\nu=1/r$ of the model. 
To gain more insight, we 
obtain analytical approximated form: 
$\Gamma/\pi \approx (1-\frac{\pi}{4})Vg^2(\omega = \frac{V}{2})
    +\frac{\pi}{4}Vg^2(\omega = 0)$ where $g(\omega)$ is 
well approximated by a semi-ellipse 
for $-V/2<\omega<V/2$ (see excellent agreement 
in Fig. 2 (c) between dotted and dashed lines)\cite{chung}.  
Via Eq.~\ref{g0gVover2} the decoherence $\Gamma$ 
at $T=0$ is approximated as: 
\begin{eqnarray}
\frac{\Gamma}{\pi V}&\approx&(1-\frac{\pi}{4})\frac{g_c^2}
{ [ 1 + ( \frac{V^2}{{T^*}^2}  \frac{\Gamma}{V} )^{-\frac{r}{2}} ]^2 } 
+ \frac{\pi}{4}\frac{g_c^2}{ [1+(\frac{V}{2T^*})^{-r}]^2 }\nonumber \\
\label{gamaoverV}
\end{eqnarray}
It is clear from Eq. ~\ref{gamaoverV} that $\Gamma/V$ 
is an universal scaling 
function of $V/T^*$. This well explains the scaling behavior obtained 
numerically (see Fig. 1 Inset). We extract further the asymptotic power-law 
behaviors of $\Gamma/V$ as a function of $V/T^*$. 
For $\Gamma \ll V\ll T^*$, we have 
$\frac{\Gamma}{\pi V}\approx \frac{\pi g_c^2}{4}(\frac{V}{2 T^*})^{2r}$. 
For $V\gg \Gamma\gg T^*$, however, we find 
$\frac{\Gamma}{\pi V}\approx \frac{\pi g_c^2}{4} [1-2(\frac{V}{2 T^*})^{-r}]$. At criticality, $\Gamma=\pi g_c^2 V$. 
The scaling behavior of the decoherence 
$\Gamma$ (Fig. 2 and Eq.~\ref{gamaoverV}), leading to distinct nonequilibrium 
scaling behaviors of all the observables discussed below, is our central 
result.\\
\emph{The conduction electron T-matrix}. 
First, we analyze nonequilibrium critical properties from 
the conduction electron T-matrix, 
defined by $G^R_{\alpha,\alpha', \sigma} = 
G^{R(0)}_{\alpha,\sigma}\delta_{\alpha, \alpha'} + G^{R(0)}_{\alpha, \sigma} 
T_{\alpha,\alpha',\sigma}(\omega) G^{R(0)}_{\alpha'\sigma}(\omega)$\cite{lars,florens} 
with $G^R_{\alpha,\alpha',\sigma}$, $G^{R(0)}_{\alpha,\alpha',\sigma}$ being 
the full and bare conduction electron Green's function, respectively. 
The imaginary part of $T-$matrix $Im[T(\omega)]\equiv T^{"}(\omega)$ 
is directly proportional to the experimentally measurable tunneling 
density of states (TDOS) of our setup. Via renormalized perturbation 
theory up to second order (see Fig. 2 (b) Inset), we have 
\begin{equation}
T^{<}_{\alpha\alpha'}(\omega) = \sum_{\beta=L,R}\int d\Omega g(\omega) 
g(\omega+\Omega) [\chi^R(\Omega) \tilde{G}^{<}_{\beta} +\chi^<(\Omega) \tilde{G}^A_{\beta}]
\end{equation}
where $\chi(\Omega)= \int_{-\infty}^{\infty} dt e^{{\it i} \Omega t} 
\chi(t)$ with $\chi(t) \equiv -\left< T_c \{\mathbf{S}^{dot}(t) \cdot 
{\mathbf{S}}^{dot}(0)\}\right>$ 
are impurity susceptibilities, $\tilde{G}^{<(A)}_{\beta}$ 
corresponds to the lesser (advanced) component of the conduction 
electron Green's functions with constant (DOS) 
(the effect of the pseudogap leads has been taken into account 
by the renormalized coupling $g(\omega)$), and 
$T^>_{\alpha\alpha'}(\omega) = T_{\alpha\alpha'}^<(-\omega)$. The imaginary 
part of the T-matrix at $T=0$ is hence given by:  
\begin{equation} 
-\pi T^{"}_{\alpha\alpha'}(\omega)=  \frac{3\pi^2}{8N(0)} g^2(\omega), 
\end{equation}
in agreement with the result in Ref.~\cite{mitra2} via a different perturbative 
RG approach to the multi-channel Kondo model out of equilibrium.
For $V=T=0$, $T^{"}(\omega)_{\alpha\alpha'}$ in the LM phase  
exhibits a power-law dip at $\omega=0$, 
$T^{"}(\omega)\propto |\omega|^{\frac{r}{2}}$. For $V>0$, 
this dip is splited into two at $\omega=\pm \frac{V}{2}$ with the same 
power-law: $T^{"}(\omega)- T^{"}(\omega=V/2) 
\propto |\omega-\frac{V}{2}|^{\frac{r}{2}}$. At the dips of $T^{"}(\omega=\pm V/2)$, 
we find $T^{"}(\omega=V/2)\propto g^2(\omega=V/2)$ 
shows a distinct nonequilibrium scaling 
behavior as a function of $V/T^*$ compared to that in equilibrium form 
$T^{"}(\omega=0)\propto g^2(\omega=0)$. To extract this different 
scaling behavior more clearly, 
we define the effective depth of the dips for 
$T^{"}(\omega=\pm V/2)$, estimated as: 
\begin{equation}
h(V)\equiv |\frac{T^{"}(\omega=V/2) - T^{"}(\omega=0)}
{T^{"}(\omega=0)}|
= |1- \frac{g^2(\omega=V/2)}{g^2(\omega=0)}|
\label{hV}
\end{equation}
It is clear from Eq. ~\ref{hV} that in the LM phase 
$h(V)$ follows an universal scaling 
function of $V/T^*$ (see Fig. 2 (b)), and has the 
following asymptotic behaviors:
$h(V) \approx 1- (\frac{4\Gamma}{V})^{r}$ for $\Gamma\ll V\ll T^*$; 
while for $T^*\ll \Gamma\ll V$, $h(V)\approx 2[(\frac{V\Gamma}{T^{*2}})^{-\frac{r}{2}}- (\frac{V}{2T^*})^{-r}]$. The 
new nonequilibrium scaling function 
$h(V)$ is detectable via STM measurement.\\
\emph{The nonequilibrium conductance}. Next, we turn our attention to the 
transport. The nonequilibrium current ${\it I}$ via the 
Fermi-Gordon rule reads\cite{Rosch,chung}:
\begin{eqnarray}
I =\frac{3\pi}{4}\int d\omega \Big[\sum_{\sigma }g_{LR}(\omega
)^{2} f_{\omega }^{L}(1-f_{\omega }^{R})\Big]-(L\leftrightarrow R).
\label{current}
\end{eqnarray}
The current ${\it I}$ is computed numerically by Eq.~\ref{current}, and is 
approximated as\cite{chung}:  
$I \approx \frac{3\pi}{4}[(1-\frac{\pi}{4})Vg^2(\omega = \frac{V}{2})
    +\frac{\pi}{4}Vg^2(\omega = 0)]$. 
The differential conductance is readily obtained numerically via 
$G=\frac{\partial {\it I}}{\partial V}$. 
In the LM phase, it has the analytical approximated form: 
\begin{eqnarray}
G(V) &\approx& 
\frac{3\pi g_c^2}{4}(1-\frac{\pi}{4})\frac{[1+(1+r)\left(\frac{V\Gamma}{{T^*}^2}\right)^{-\frac{r}{2}}]}{\left[1+\left(\frac{V\Gamma}{{T^*}^2}\right)^{-\frac{r}{2}}\right]^3}
\nonumber \\ 
&+&\frac{3\pi^2 g_c^2}{16}\frac{[1+(1+2r)\left(\frac{V}{2T^*}\right)^{-r}]}{\left[1+\left(\frac{V}{2T^*}\right)^{-r}\right]^3}
\label{GV-approx}
\end{eqnarray}
As shown in Fig. 2 (c), for $T^*\ll V\ll D_0$, $G(V)$ 
approaches the equilibrium 
scaling form
\begin{equation} 
G^{eq}(V\rightarrow T)\approx \frac{3\pi}{4} g^{eq}(T)^2 = 
\frac{\frac{3\pi}{4}g_c^2}{[1+ (T/T^{\ast})^{-r}]^2}
\label{GV-eq}
\end{equation}
; while as for $V/T^*\gg 1$ 
it exhibits a distinct 
universal scaling behavior of $V/T^*$. 
The perfect scaling behavior of $G(V/T^*)$ is a direct consequence  
of the $V/T^*$ scaling in $\Gamma/V$.   
By contrast, the universal $V/T^*$ scaling is absent 
in Ref.\cite{chung} as $\Gamma /V$ is not a universal 
function of $V/T^*$ there. 
For $\Gamma\ll V \ll T^*$, the conductance behaves as: 
$G(V)\approx [a (\frac{V}{2T^*})^{2r} + 
b (\frac{V}{T^*})^{2r+2r^2}]$ with $a = \frac{3\pi^2 r^2}{64}$, 
$b = \frac{3\pi r^2}{16} (1-\frac{\pi}{4}) (\frac{r^2}{8})^{r}$, 
which shows a prefactor reduction in the first term $\propto V^{2r}$ 
with respect to its equilibrium form $G^{eq}(V\rightarrow T\ll T^*)\approx 
\frac{3\pi r^2}{16} (\frac{T}{T^*})^{2r}$ and 
a sub-leading correction with an anomalous power-law 
behavior $\propto V^{2r+2r^2}$. 
For $V\gg \Gamma \gg T^*$, however, 
we find $G(V) \approx 
\frac{3\pi^2 r^2}{64} [1- p (\frac{V}{2T^*})^{-r} -q 
(\frac{V}{T^*})^{-2r}]$ with $p= 1-r$ and $q = \frac{8}{\pi} 
(1-\frac{\pi}{4}) (1-r) (\frac{r^2}{8})^{-\frac{r}{2}}$, which 
deviates significantly from its equilibrium form 
$G^{eq}(V\rightarrow T\gg T^*)\approx \frac{3\pi r^2}{16} 
[1- 2 (\frac{T}{T^*})^{-r}]$.  
It is worthwhile emphasizing that due to the very different role played 
by the bias $V$ and temperature $T$, 
$G(V)$ follows a completely different scaling function from its equilibrium 
form $G^{eq}(V\rightarrow T)$ over the full range of $V/T^*$ 
(see Eq.~\ref{GV-approx} and Eq. ~\ref{GV-eq}) 
though it tends to converge with its equilibrium form for $V\ll T^*$.\\
\emph{Local spin susceptibility.}
We furthermore analyze the scaling 
behaviors of the local spin susceptibility $\chi_{loc}(V)\equiv 
\frac{\partial M}{\partial h}|_{h\rightarrow 0}$ 
in the LM phase with $h$ being a small magnetic field, 
$M$ being the magnetization 
$M=\frac{n_{\uparrow}-n_{\downarrow}}{n_{\uparrow}+n_{\downarrow}}$.   
Following Ref.~\cite{Rosch,chung,chung2}, 
for $V\gg h\rightarrow 0$, we find  
$M\approx \frac{\int_{\frac{-V-h}{2}}^{\frac{-V+h}{2}} d\omega g^2(\omega)}{ 
\int_{\frac{-V}{2}}^{\frac{V}{2}} d\omega g^2(\omega)}$. The 
approximated form for $V\chi_{loc}$ reads (see Fig. 2 (d))\cite{latha}: 
$V\chi_{loc} \approx 
\frac{g(\omega=V/2)^2}{\frac{\pi}{4} g(\omega=0)^2 +(1-\frac{\pi}{4}) 
g(\omega=V/2)^2}$. 
For $\Gamma\ll V \ll T^*$, $\chi_{loc}$ exhibits an anomalous 
power-law behavior: 
$\chi_{loc} \propto \frac{1}{V^{1-\eta_{\chi}}}$ with $\eta_{\chi}= 2r^2$, 
distinct from its equilibrium constant behavior: 
$T\chi_{loc}(T\ll T^*)\propto (g_c-g)^r$\cite{florens}. 
For $V\gg \Gamma \gg T^*$, however, we find the 
local susceptibility acquires a power-law correction to the Curie behavior:  
$V\chi_{loc} \approx 1- V\Delta\chi_{loc}$ and $\Delta\chi_{loc}\propto \frac{1}{V^{1-\Delta\eta_{\chi}}}$ 
with an anomalous exponent $\Delta\eta_{\chi}=-r$; 
while as its corresponding equilibrium form 
shows a different anomalous power-law 
behavior: $\chi_{loc}(T\gg T^*)\propto \frac{1}{T^{1-\eta_{\chi}}}$ with 
$\eta_{\chi}= r^2$\cite{lars,florens}. At criticality ($g=g_c$), 
$\chi_{loc}$ shows perfect Curie law behavior: $\chi_{loc}\propto 1/V$. 
These distinct nonequilibrium signatures near QCP 
are detectable in local susceptibility measurement.\\
\emph{Conclusions.} In summary, via a controlled frequency-dependent renormalization group approach we have investigated the quantum phase transition out of equilibrium in the pseudogap Kondo quantum dot. At zero temperature and finite bias voltage, we discovered in the local moment phase the new quantum critical behaviors in the T-matrix, conductance, and local spin susceptibility that are distinct from those in equilibrium and at finite temperatures. The key to explain these differences lies in the fact that the current-induced decoherence at 
a finite bias voltage (out of equilibrium) acts quite differently from that at a finite temperature but zero bias (in equilibrium), 
resulting in distinct nonequilibrium behavior near the quantum phase transition. Our predictions open up a new perspective both theoretically and experimentally in the study of the Kondo dot coupled to exotic leads with pseudogap density of states.



\acknowledgements

We thank  M. Vojta for many helpful discussions.
This work is supported by the NSC grant 
No.98-2112-M-009-010-MY3, the MOE-ATU program, the
NCTS of Taiwan, R.O.C.         .



\vspace*{-10pt}
\begin{thebibliography}{}
\vspace*{-10pt}

\bibitem{subir}
S. Sachdev, {\it Quantum Phase Transitions}, Cambridge University Press (2000); 
S. L. Sondhi, S. M. Girvin, J. P. Carini, and D. Shahar,
Rev. Mod. Phys. {\bf 69}, 315 (1987).


\bibitem{Goldhaber}
R. M. Potok, I. G. Rau, H. Shtrikman, Y. Oreg  and D. Goldhaber-Gordon, 
 Nature \textbf{447} 167-171 (2007).

\bibitem{Hewson}
A.C. Hewson, The Kondo Problem to Heavy Fermions, Cambridge University 
Press, Cambridge (1997).

\bibitem{NoneqQPT}
D. E. Feldman, Phys. Rev. Lett., {\bf 95}, 177201 (2005); 
A. Mitra, S. Takei, Y.B. Kim, and A. J. Millis,
Phys. Rev. Lett., {\bf 97}, 236808 (2006); 
S. Takei, Y.B. Kim, Phys. Rev. B {\bf 76} 115304 (2007); 
S. Kirchner, Q.M. Si, Phys. Rev. Lett. {\bf 103}, 206401 (2009).

\bibitem{chung}
C.-H. Chung, K. Le Hur, M. Vojta and P. W\"{o}lfle,
Phys. Rev. Lett. \textbf{102}, 2106803 (2009).

\bibitem{chung2}
C.H. Chung, K.V.P. Latha, K. Le Hur, M. Vojta and P. W\"olfle, 
 Phys. Rev. B, {\bf 82}, 115325 (2010). 


\bibitem{fradkin}
D. Withoff, E. Fradkin, Phys. Rev. Lett. {\bf 64}, 1835 (1990).

\bibitem{GBI}
C.\ Gonzalez-Buxton and K.\ Ingersent, \prb \textbf{57}, 14254 (1998).

\bibitem{insi}
K.\ Ingersent and Q.\ Si, Phys.\ Rev.\ Lett.\ \textbf{89}, 076403 (2002).



\bibitem{lars}
M.\ Vojta and L.\ Fritz, Phys.\ Rev.\ B \textbf{70}, 094502 (2004); 
L.\ Fritz and M.\ Vojta, Phys.\ Rev.\ B \textbf{70}, 214427 (2004).

\bibitem{florens}
Lars Fritz, Serge Florens, Matthias Vojta, Phys. Rev. {\bf B} 
74, 144410 (2006). 

\bibitem{hurpsg}
John Hopkinson, Karyn Le Hur, Emilie Dupont, Physica B, {\bf 359-361} 1454 (2005).

\bibitem{vojtagraphene}
Matthias Vojta, Lars Fritz, Ralf Bulla, Eur. Phys. Lett. {\bf 90}, 27006 (2010). 

\bibitem{nancy}
 Luis G. G. V. Dias da Silva, Nancy Sandler, Pascal Simon, Kevin Ingersent, Sergio E. Ulloa, 
Phys. Rev. Lett. {\bf 102} 166806 (2009).  

\bibitem{trueQPT}
The crossover scale $T^*$ associated with the ``true'' QPT is a power-law 
function of the distance $t=|g-g_c|$ to QCP; while as it depends exponentially on $t$ for QPT of the KT type.


\bibitem{Rosch}
A. Rosch, J. Paaske, J. Kroha, P. W\"{o}lfle, Phys. Rev. Lett. \textbf{90}, 076804 (2003); J. Phys. Soc. Jpn. \textbf{74}, 118 (2005).





\bibitem{latha}
J. Paaske, A. Rosch, Phys. Rev. B {\bf 69} 155330 (2004); 
Chung-Hou Chung, K.V.P. Latha, Phys. Rev. {\bf B} 82, 085120 (2010).


\bibitem{fRG}
S. Kehrein, Phys. Rev. Lett. \textbf{95}, 056602 (2005); 
H. Schoeller, F. Reininghaus, Phys. Rev. B {\bf 80}, 045117 (2009); 
S.G. Jakobs, V. Meden and H. Schoeller, Phys. Rev. Lett. {bf 99}, 150603 (2007).


\bibitem{woelflefRG}
H. Schmidt and P. W\"oelfle, 
Ann. Phys. (Berlin) {\bf 19}, No. 1-2, 60-74 (2010). 

\bibitem{mitra2}

A. Mitra, A. Rosch, Phys. Rev. Lett. {\bf 106} 106402, (2011).

\end{thebibliography}
\end{document}